\documentclass[aps,prd,showpacs,nofootinbib,tightenlines]{revtex4}
\usepackage{amsmath,amsthm,amsmath}
\usepackage{amssymb}
\usepackage{epsfig}
\usepackage{graphicx}
\usepackage{bm}
\usepackage{times}
\usepackage{braket}
\usepackage{color}
\usepackage{slashed}
\usepackage{hyperref}
\usepackage{epsfig}
\usepackage{url}
\usepackage[normalem]{ulem}
\usepackage[utf8]{inputenc}
\begin{document}
%%%%%%%%%%%%%%%%%%%%%%%%%%%%%%%%%%%%%%%%
\title{Resonance contributions from $\chi_{c0}$ in the charmless three-body hadronic $B$ meson decays}
\vspace{1.0cm}
\author{Ya-Lan Zhang$^{1}$}
\email{zylyw@hyit.edu.cn}
\author{Chao Wang$^{1}$} 
\email{chaowang@nankai.edu.cn}
\author{Yi Lin$^{1}$}
 \email{linyihit@hyit.edu.cn}
\author{Zhen-Jun Xiao$^{2,3}$} 
\email{xiaozhenjun@njnu.edu.cn}
\affiliation{$^1$Faculty of Mathematics and Physics, Huaiyin Institute of Technology, Huaian, Jiangsu 223200, P.R. China}
\affiliation{$^2$Department of Physics and Institute of Theoretical Physics, Nanjing Normal University, Nanjing, Jiangsu 210023, P.R. China}
\affiliation{$^3$Jiangsu Key Laboratory for Numerical Simulation of Large Scale Complex Systems, Nanjing Normal University, Nanjing, Jiangsu 210023, P.R. China}
\date{\today}
%%%%%%%%%%%%%%%%%%%%%%%%%%%%%%%%%%%%%%%%

\begin{abstract}
{\bf Abstract:}Within the framework of perturbative QCD factorization, we investigate the nonfactorizable contributions to these factorization-forbidden Quasi-two-body decays $B_{(s)}\rightarrow h\chi_{c0}\rightarrow h\pi^+\pi^-(K^+K^-)$ with $ h=\pi, K $.
We compare our predicted branching ratios for the $B_{(s)}\rightarrow K\chi_{c0}\rightarrow K\pi^+\pi^-(K^+K^-)$ decay with available experiment data
as well as predictions by other theoretical studies.
The branching ratios of these decays are consistent with data and other theoretical predictions.
In the Cabibbo-suppressed decays $B_{(s)}\rightarrow h\chi_{c0}\rightarrow h\pi^+\pi^-(K^+K^-)$ with $h=\bar{K}^0,\pi$, however, the values of the branching ratios are the order of $10^{-7}$ and $10^{-8}$.
The ratio $R_{\chi_{c0}}$ between the decay $B^+\rightarrow \pi^+\chi_{c0}\rightarrow \pi^+\pi^+\pi^-$ and $B^+\rightarrow K^+\chi_{c0}\rightarrow K^+\pi^+\pi^-$ and the distribution of branching ratios for different decay modes in invariant mass are considered in this work.

\end{abstract}


%%%%%%%%%%%%%%%%%%%%%%%%%%%%%%%%%%%%%%%%
\footnote{Supported by the National Natural Science Foundation of China (11235005, 11847141, 12105112), the Natural Science Foundation of Jiangsu Education Committee (21KJB140027) and the Natural Science Foundation of Huaiyin Institute of Technology (21HGZ012).}
%%%%%%%%%%%%%%%%%%%%%%%%%%%%%%%%%%%%%%%%

\maketitle
%%%%%%%%%%%%%%%%%%%%%%%%%%%%%%%%%%%%%%%%
\section{INTRODUCTION}
The $P$-wave $0^{++}$ charmonium state $\chi_{c0}$ cannot be created by the colorless current $\bar c\gamma^\mu(1-\gamma_5)c$.
Its production in the $B$ meson decays is suppressed in the factorization approximation because of the charge conjugation
invariance, spin-parity and vector current conservation~\cite{zpc34103,prd59054003,prd66037503,jhep0106067}.
Thus, it's not surprising that the report for the branching fraction $\mathcal{B}(B^+\to K^+\chi_{c0})=(6.0^{+2.1}_{-1.8} ({\rm stat.}) \pm1.1({\rm syst.}))\times10^{-4}$~\cite{prl88031802} from Belle Collaboration triggered many studies on this two-body decay mode involving
$\chi_{c0}$ and the relevant decay processes.
The measurement by  BaBar Collaboration in 2004 confirmed Belle's result for $B^+\to K^+\chi_{c0}$ and presented the value $(2.7\pm0.7)\times10^{-4}$ for its branching fraction~\cite{prd69071103}.
The recent data in {\it Review of Particle Physics} for this two-body decay process is $1.51^{+0.15}_{-0.13}\times 10^{-4}$~\cite{PDG2022}, which is only about $1/4$ of its first appearance~\cite{prl88031802} but still some comparable to that of the factorization-allowed decay $B^+\to K^+J/\psi$ which has the branching fraction $(1.020\pm0.019)\times 10^{-3}$ in~\cite{PDG2022}.

Since Belle's measurement~\cite{prl88031802}, numerous theoretical studies have been conducted to investigate the large nonfactorizable contributions, the decay characteristic in $B^+\to K^+\chi_{c0}$ and other relevant decay modes.
In the light-cone QCD sum rules approach, the nonfactorizable soft contributions in the $B\to K\eta_c, K\chi_{c0}$ decays were analyzed in the
Ref.~\cite{prd70074006}.
Within the perturbative QCD (PQCD) approach, the nonfactorizable contributions to the $B$ meson decays into charmonia including $B^{0,+}\to K^{(\ast)0,+} \chi_{c0}$ were calculated in the Refs.~\cite{prd71114008,cpc44113104}.
In the framework of QCD factorization (QCDF), the exclusive decays including the $B \to \chi_{c0}K$ were studied
in~\cite{prd69054009,plb568127,ctp48885,0607221,plb619313,npb811155}.
From these studies it was observed that infrared divergences resulting from nonfactorizable vertex corrections could not be eliminated~\cite{prd69054009,plb568127}. Non-zero gluon mass was then employed to regularize the infrared divergences in vertex corrections~\cite{ctp48885}.
While the authors of~\cite{npb811155} found those infrared divergences can be subtracted consistently into the matrix elements of colour-octet operators in the exclusive $B$ to $P$-wave charmonia decays.
In Ref.~\cite{plb619313}, the $B\to K\chi_{c0,2}$ decays were investigated in QCDF by introducing a non-zero binding energy to regularize the infrared divergence of the vertex part and adopting a model dependent parametrization to remove the logarithmic and linear infrared divergences in the spectator diagrams.
The rescattering effects mediated by intermediate charmed mesons were studied in Refs.~\cite{plb54271,epjc33S247}, the authors concluded that such effects could produce a large branching ratio for the decay $B^+\to K^+\chi_{c0}$.

Unlike the $2P$ state $\chi_{c0}^\prime$, which will mainly decay to $D\bar D$ in an $S$-wave~\cite{prd95112003,prd86091501,rmp90015003}, the state $\chi_{c0}$, with its mass below the threshold of $D\bar D$~\cite{PDG2022}, can decay into light hadronic states via gluon-rich processes~\cite{pr411,prd72074001,plb659221}.
Although the branching fractions for $\chi_{c0}\to\pi^+\pi^-$ and $\chi_{c0}\to K^+K^-$ are small, in the order of $10^{-3}$~\cite{PDG2022}, the resonance contributions from $\chi_{c0}$ are not negligible in the three-body decays $B\to h\pi^+\pi^- (K^+K^-)$ because of the enhancements originate from the Cabibbo-Kobayashi-Maskawa (CKM) matrix elements when compared with the resonant states $\rho(770)$ or $\phi(1020)$\cite{prd103056021,prd101119901},
where $h$ is a light pseudoscalar $\pi$ and $K$, $B$ meson is $B^{+,0}$ and $B^0_s$, and the inclusion of charge-conjugate processes are implied throughout this work.
Taking the $B^+\to K^+K^-K^-$ as an example, the branching fraction is $(1.12\pm0.15\pm0.06)\times10^{-6}$ for the quasi-two-body
decay process $B^+\to K^+\chi_{c0}\to K^+K^+K^-$ in~\cite{prd85112010}, which is about $1/4$ of the process $B^+\to K^+\phi(1020)\to K^+K^+K^-$ and is about $3.24\%$ of the total branching fraction for $B^+\to K^+K^+K^-$~\cite{prd85112010}.
The fit fraction for the quasi-two-body $B^+\to K^+\chi_{c0}\to K^+\pi^+\pi^-$ is $(1.12\pm0.12^{+0.24}_{-0.08})\%$ in~\cite{prl96251803} and $(3.56\pm0.93)\%$ (the model $A_0$) in~\cite{prd71092003}, respectively.
So, it is important to study the resonance contributions from $\chi_{c0}$ in the charmless three-body hadronic $B$
meson decays, and this research will improve a comprehension understanding for three-body decay.

In this work, we will systematically analyze the contributions from $\chi_{c0}$ in the decays $B\to h\pi^+\pi^- (K^+K^-)$ in the PQCD
approach~\cite{plb5046,prd63054008,prd63074009,ppnp5185}, which has been adopted to study the three-body $B$ meson
decays~\cite{plb561258,prd70054006,prd89074031,prd91094024,prd97034033}.
With the help of the experimental inputs for the time-like pion form factors~\cite{prd86032013} and the two-pion distribution amplitudes~\cite{fp42101,prl811782,npb555231}, the decays $B\to K\rho(770), K\rho^\prime(1450)\to K\pi\pi$\cite{plb76329}, $B\to K^{\ast}_0(1430)h, K^{\ast}_0(1950)h \to K\pi h$ \cite{jhep03162}, $B\to D^{\ast}(2007)^0h, D^{\ast}(2010)^{\pm}h\to D\pi h$\cite{plb791342} and $B^0_{(s)}\to \eta_c(2s) \pi^+\pi^-$ \cite{cpc41083105} were analyzed in the quasi-two-body framework.
The method used in~\cite{plb76329} have been adopted for other quasi-two-body $B$ meson decays in the Refs.~\cite{prd104116019,cpc46053104,prd103016002,prd103096016,epjc7937,prd100014017,plb788468} in recent years.
For the detailed discussions of the quasi-two-body framework based on the PQCD approach, we refer to the Refs.~\cite{plb76329,plb788468}.

%%%%%%%%%%%%%%%%%%%%%%%%%%%%%%%%%%%%%%%%%%%%%%
\begin{figure}[!th] %htbp
\begin{center}
\includegraphics[width=0.6\textwidth]{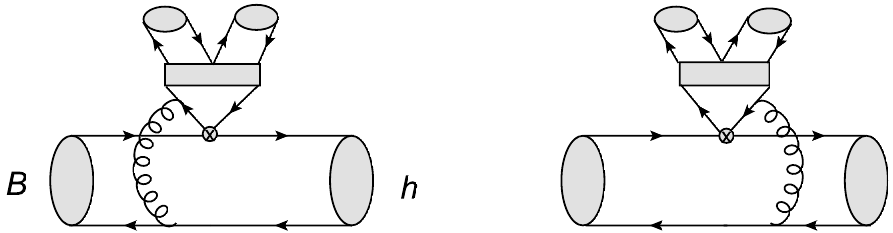}
\caption{Feynman diagrams for spectator figure to $B_{(s)}\rightarrow h \chi_{c0}\rightarrow h\pi^+\pi^-$ and
               $B_{(s)}\rightarrow h \chi_{c0}\rightarrow hK^+K^-$.}
\vspace{-0.5 cm}
\label{fig:fig1}
\end{center}
\end{figure}
%%%%%%%%%%%%%%%%%%%%%%%%%%%%%%%%%%%%%%%%%%%%%%%
\section{FRAMEWORK}
Under the factorization hypothesis, the decay amplitude for $B\to h \chi_{c0} \to hK^+K^-$ is given by
\begin{eqnarray}
\langle h(K^+K^-)_{\chi_{c0}} | \mathcal{H}_{\rm eff}| B \rangle &\simeq& \langle K^+K^-|\chi_{c0}\rangle\frac{1}{\mathcal{D}_{\rm BW}}
\langle h\chi_{c0} | \mathcal{H}_{\rm eff}| B \rangle \nonumber \\
    &=&\frac{g_{\chi_{c0}K^+K^-}}{\mathcal{D}_{\rm BW}}\langle h\chi_{c0} | \mathcal{H}_{\rm eff}| B \rangle \nonumber \\
    &=& C_{KK}(s)\cdot \mathcal{A}(s)\;,
\label{eq-DA-KK}
\end{eqnarray}
where the denominator ${\mathcal{D}_{\rm BW}}=m^2_0-s-im_0\Gamma(s)$, the mass-dependent decay width $\Gamma(s)$ is defined as $\Gamma(s)=\Gamma_0\frac{m_0}{\sqrt s}(\frac{q}{q_0})^{2L_R+1}$, $m_0=(3414.71\pm0.30)$ \,{\rm MeV} and
$\Gamma_0=(10.8\pm0.6)$ \,{\rm MeV}~\cite{PDG2022} are the pole mass and full width of the resonant state $\chi_{c0}$, the $s$ is
invariant mass square for $K^+K^-$ pair in the decay final state.
$L_R$ is the spin of the resonances\cite{prd85112010,prd71092003}.
In the rest frame of the resonant state $\chi_{c0}$, its daughter $K^+$ or $K^-$ has the magnitude of the momentum as {\small $q=\frac{1}{2}\sqrt{s-4m_K^2}$}, and $q_0$ in $\Gamma(s)$ is the value of $q$ at $s=m^2_0$.
The amplitude $\mathcal{A}(s)=\langle h\chi_{c0} | \mathcal{H}_{\rm eff}| B \rangle$ for the
concerned quasi-two-body decays in this work can be found in the Appendix.
The mass-dependent coefficient $C_{KK}(s)$ is $g_{\chi_{c0}K^+K^-}/\mathcal{D}_{\rm BW}$.
We have the coupling constant $g_{\chi_{c0}K^+K^-}$ from the relation~\cite{prd92014031,prd68094005}
\begin{eqnarray}
g_{\chi_{c0}K^+K^-}=\sqrt{\frac{8\pi m^2_0 \Gamma_{\chi_{c0}\to K^+K^-}}{q_0}}\;,
\label{eq-gKK}
\end{eqnarray}
where the $\Gamma_{\chi_{c0}\to K^+K^-}$ is the partial width for $\chi_{c0}\to K^+K^-$.
For the process  $B\to h \chi_{c0} \to h\pi^+\pi^-$, we need the replacement $K\to \pi$ for the Eqs.~(\ref{eq-DA-KK})-(\ref{eq-gKK})
and the relevant parameters.
The effective Hamiltonian $\mathcal{H}_{\rm eff}$ with the four-fermion operators are the same as in~\cite{prd71114008}.

In the rest frame of the $B$ meson, we choose its momentum $p_B$, the momenta $p_3$ and $p$ for the bachelor state $h$
and $\chi_{c0}$, as
\begin{eqnarray}
p_B&=&\frac{m_B}{\sqrt{2}}(1,1,{\textbf 0_{\rm T}}), \quad\quad\;\;
p_3=\frac{m_B}{\sqrt{2}}(0,1-\eta,{\textbf 0_{\rm T}}), \quad\quad\quad
p=\frac{m_B}{\sqrt{2}}(1,\eta,{\textbf 0_{\rm T}}),\nonumber\\
k_B&=&(0,\frac{m_B}{\sqrt{2}}x_B,k_{\rm BT}),\quad
k_3=(0,\frac{m_B}{\sqrt{2}}(1-\eta)x_3,k_{\rm 3T}),\quad
k=(\frac{m_B}{\sqrt{2}}z,\frac{m_B}{\sqrt{2}}z\eta,k_{\rm T}),
\label{eq:m}
\end{eqnarray}
where $x_B$, $x_3$, and $z$ are the corresponding momentum fractions, $m_B$ is the mass of $B$ meson.
The variable $\eta$ is defined as $\eta=s/m_B^2$, with the invariant mass square $s=p^2$.
For the $B^{+,0}$ and $B^0_s$ in this work, we employ the same distribution amplitudes $\phi_{B/B_s}$ as in Refs.~\cite{prd86114025,prd89074031}.
The wave functions for the bachelor states $\pi$ and $K$ in this work are written as
\begin{eqnarray}
\Phi_{h}(p,z)=\frac{1}{\sqrt{2N_c}}\gamma_5(p\hspace{-1.5truemm}/\phi^{A}(z)+ m_0^h\phi^{P}(z)+ m_0^h(n\hspace{-2truemm}/v\hspace{-2truemm}/-1)\phi^{T}(z)),
\label{eq:b}
\end{eqnarray}
where $m_0^h$ is the chiral mass, $p$ and $z$ are the momentum and corresponding momentum fraction of $\pi$ and $k$.
The distribution amplitudes (DAs) $\phi^{A}(z)$, $\phi^{P}(z)$, $\phi^{T}(z)$ can be written as\cite{jhep09005,jhep01010,prd71014015,jhep05004}
\begin{eqnarray}
\phi^{A}(z)&=&\frac{f_h}{2\sqrt{2N_c}}6z(1-z)[1+a_1^hC_1^{3/2}(t)+a_2^hC_2^{3/2}(t)+a_4^hC_4^{3/2}(t)],\nonumber\\
\phi^{P}(z)&=&\frac{f_h}{2\sqrt{2N_c}}[1+(30\eta_3-\frac{5}{2}\rho_h^2)C_1^{1/2}(t)-3[\eta_3\omega_3+\frac{9}{20}\rho_h^2(1+6a_2^h)]C_4^{1/2}(t)],\nonumber\\
\phi^{T}(z)&=&\frac{f_h}{2\sqrt{2N_c}}(1-2z)[1+6(5\eta_3-\frac{1}{2}\eta_3\omega_3-\frac{7}{20}\rho_h^2-\frac{3}{5}\rho_h^2a_2^h)(1-10z+10z^2)],
\label{eq:APT}
\end{eqnarray}
where the Gegenbauer moments are chosen as $a_1^{\pi}=0$, $a_1^K=0.06$, $a_2^{\pi,K}=0.25\pm0.15$, $a_4^{\pi}=-0.015$ and the paraments follow
$\rho_{\pi}= m_{\pi}/m_{0}^{\pi}$, $\rho_{K}= m_{K}/m_{0}^{K}$, $\eta_3^{\pi,K}=0.015$, $\omega_3^{\pi,K}=-3$.
We adopt $m_{0}^{\pi}=(1.4\pm0.1)\rm{GeV}$, $m_{0}^{K}=(1.6\pm0.1)\rm{GeV}$ in the numerical calculations.
The Gegenbauer polynomials are defined as
\begin{eqnarray}
C_1^{\frac{3}{2}}(t)&=&3t,\quad C_2^{\frac{1}{2}}(t)=\frac{1}{2}(3t^{2}-1),\quad C_2^{\frac{3}{2}}(t)=\frac{3}{2}(5t^{2}-1),\nonumber\\
C_4^{\frac{1}{2}}(t)&=&\frac{1}{8}(3-30t^2+35t^4),\quad C_4^{\frac{3}{2}}(t)=\frac{15}{8}(3-30t^2+35t^4),
\label{eq:Gegenbauer polynomials}
\end{eqnarray}
where the variable $t=2z-1$.
The mass-dependent $\pi\pi$ or $KK$ system, which comes from $\chi_{c0}$, has the distribution amplitude~\cite{prd71114008}
\begin{eqnarray}
\Phi_{\pi\pi(KK)}=\frac{1}{\sqrt{2N_c}}(p\hspace{-1.5truemm}/\phi_{\pi\pi(KK)}^v(z)+\sqrt{s}\phi_{\pi\pi(KK)}^s(z)),
\label{eq:P}
\end{eqnarray}
with the twist-2 and twist-3 distribution amplitudes $\phi_{\pi\pi(KK)}^v(z,s)$ and $\phi_{\pi\pi(KK)}^s(z,s)$
\begin{eqnarray}
\phi_{\pi\pi(KK)}^v(z,s)&=&\frac{F_{\chi_{c0}}(s)}{2\sqrt{2N_c}}27.46(1-2z)\left\{\frac{z(1-z)[1-4z(1-z)]}{[1-2.8z(1-z)]^2}\right\}^{0.7},\nonumber\\
\phi_{\pi\pi(KK)}^s(z,s)&=&\frac{F_{\chi_{c0}}(s)}{2\sqrt{2N_c}}4.73\left\{\frac{z(1-z)[1-4z(1-z)]}{[1-2.8z(1-z)]^2}\right\}^{0.7}.
\label{eq:pvs}
\end{eqnarray}
The timelike form factor $F_{\chi_{c0}}(z,s)$ is parametrized with the RBW line shape\cite{prl120261801} and can be expressed as follows\cite{prd101016015,epjc80394,cpc46123105},
\begin{eqnarray}
F_{\chi_{c0}}(s)=\frac{m^2_{0}}{m^2_{0}-s-im_{0}\Gamma_{(s)}},
\label{eq:fp}
\end{eqnarray}
where $m_{0}$ is the pole mass. The mass-dependent decay width $\Gamma_{(s)}$ is defined as
\begin{eqnarray}
\Gamma(s)=\Gamma_0\frac{m_0}{\sqrt s}(\frac{q}{q_0})^{2L_R+1},
\label{eq:gs}
\end{eqnarray}
$L_R$ is the spin of the resonances, and $L_R=0$ for the scalar intermediate state $\chi_{c0}$.

\section{RESULTS}
The differential branching ratios ($\mathcal B$) for the decay processes $B\to h\pi^+\pi^-(K^+K^-)$ is
\begin{eqnarray}
\frac{d\mathcal{B}}{ds}=\tau_B\frac{q_h q }{64\pi^3m^3_B }{\overline{|C_{\pi\pi(KK)}\cdot\mathcal{A}}|^2},
\label{eq:B}
\end{eqnarray}
where $\tau_B$ is the lifetime of $B$ meson.
The $q_h$ is the magnitude momentum for the bachelor $h$ in the rest frame of $\chi_{c0}$:
\begin{eqnarray}
q_h=\frac{1}{2}\sqrt{\left[ (m_B^2-m_h^2)^2-2(m_B^2+m_h^2)s+s^2 \right]/s},
\label{eq:qh}
\end{eqnarray}
with $m_h$ is the mass of $h$.
The central values (in units of GeV) of the relevant mesons and quark masses are adopted as\cite{PDG2022}
\begin{eqnarray}
m_{B}=5.279, \quad m_{B_s}=5.367,\quad
m_{\pi^{\pm}}=0.140,\quad m_{\pi^0}=0.135,\nonumber\\
m_{K^{\pm}}=0.494,\quad m_{K^0}=0.498,\quad
m_{b}(pole)=4.8,\quad m_{c}=1.275.
\label{eq:mass}
\end{eqnarray}
For the decay constants(in units of GeV) and lifetimes(in units of ps) of the relevant mesons, we use\cite{PDG2022}
\begin{eqnarray}
f_{B}=0.19, \quad f_{B_s}=0.227,\quad f_{\chi_{c0}}=0.36,
\quad f_{\pi}=0.131,\nonumber\\
\quad f_{K}=0.156,
\quad\tau_{B^\pm}=1.638,\quad \tau_{B^0}=1.52,\quad \tau_{B_s}=1.51.
\label{eq:dc}
\end{eqnarray}
The QCD scale follows $\Lambda_{\overline{MS}}^{(f=4)}=0.25{\rm GeV}$.
We adopt the Wolfenstein parameters$(A,\overline{\lambda},\overline{\rho},\overline{\eta})$ of CKM mixing matrix $A=0.836\pm0.015$, $\overline{\lambda}=0.22453\pm0.00044$, $\overline{\rho}=0.122^{+0.018}_{-0.017}$, $\overline{\eta}=0.335^{+0.012}_{-0.011}$\cite{PDG2022}.
For the shape parameter uncertainty of $B_{(s)}$ meson we use $\omega_B=0.4\pm0.04\,{\rm GeV}$ and $\omega_{B_s}=0.5\pm0.05\,{\rm GeV}$, which contributed the largest error for the branching fractions.
The second one is from the Gegenbauer moments $a_2^{h}$ in the bachelor meson DAs.
The other two error comes from decay width of the resonance $\chi_{c0}$ and the chiral mass $m_0^{h}$ of bachelor meson, which have a smaller impact to the uncertainties in our approach.
There are further errors which are tiny and can be ignored safely, such as minor and disregarded parameters in the bachelor meson $(\pi/K)$ distribution amplitudes and Wolfenstein parameters.

\begin{table}[thb]
\begin{center}
\caption{PQCD predictions of branching ratios for the quasi-two-body decays $B_{(s)}\rightarrow h\chi_{c0}\rightarrow h\pi^+\pi^-(K^+K^-)$.}
\label{1}
\vspace{0cm}
\begin{tabular}{l  c c c c} \hline \hline
Mode~~~~~~~~~~~~~~~~~~&~~~~~~~~~~~~~~~~~~Unit~~~~~~~~~~~~~~~~~~&~~~~~~~~~~~~~~~~~~Branching ratios~~~~~~~~~~&~~~~~~~~~~~~~~Data\cite{PDG2022}~~~~~~~~~~~~~~  \\
\hline
       $B^+\rightarrow K^+\chi_{c0}\rightarrow K^+\pi^+\pi^-$ & $(10^{-6})$ &$0.81_{-0.22}^{+0.21}(\omega_B)_{-0.21}^{+0.12}(a_2)_{-0.05}^{+0.11}(\Gamma_{\chi_{c0}})_{-0.06}^{+0.06}(m_0^K)$&$-$\\
       $B^+\rightarrow K^+\chi_{c0}\rightarrow K^+K^+K^-$ & $(10^{-6})$ &$0.84_{-0.24}^{+0.22}(\omega_B)_{-0.12}^{+0.08}(a_2)_{-0.08}^{+0.07}(\Gamma_{\chi_{c0}})_{-0.02}^{+0.02}(m_0^K)$&$-$\\
%       $0.7\pm0.10_{-0.10}^{+0.12}$ \cite{prd78-012004}\\

       $B^0\rightarrow K^0\chi_{c0}\rightarrow K^0\pi^+\pi^-$ & $(10^{-6})$
       &$1.21_{-0.32}^{+0.55}(\omega_B)_{-0.13}^{+0.27}(a_2)_{-0.10}^{+0.12}(\Gamma_{\chi_{c0}})_{-0.05}^{+0.01}(m_0^K)$&$-$\\
       $B^0\rightarrow K^0\chi_{c0}\rightarrow K^0 K^+K^-$ & $(10^{-6})$ &$1.30_{-0.27}^{+0.32}(\omega_B)_{-0.16}^{+0.22}(a_2)_{-0.02}^{+0.01}(\Gamma_{\chi_{c0}})_{-0.04}^{+0.01}(m_0^K)$&$-$\\
%       $0.496\pm0.02$\cite{prl120-261801} \\
       \hline
       $B_s^0\rightarrow \bar{K}^0\chi_{c0}\rightarrow \bar{K}^0\pi^+\pi^-$ & $(10^{-7})$
       &$1.86_{-0.28}^{+0.41}(\omega_B)_{-0.22}^{+0.38}(a_2)_{-0.12}^{+0.14}(\Gamma_{\chi_{c0}})_{-0.04}^{+0.03}(m_0^K)$&-\\
       $B_s^0\rightarrow \bar{K}^0\chi_{c0}\rightarrow \bar{K}^0 K^+K^-$ & $(10^{-7})$
       &$2.45_{-0.61}^{+0.33}(\omega_B)_{-0.26}^{+0.39}(a_2)_{-0.23}^{+0.45}(\Gamma_{\chi_{c0}})_{-0.02}^{+0.02}(m_0^K)$&-\\
       \hline
       $B^+\rightarrow \pi^+\chi_{c0}\rightarrow \pi^+\pi^+\pi^-$ &$ (10^{-8})$ &$3.93_{-0.70}^{+0.65}(\omega_B)_{-0.54}^{+0.40}(a_2)_{-0.21}^{+0.18}(\Gamma_{\chi_{c0}})_{-0.01}^{+0.01}(m_0^{\pi})$&  $<10$ \\
       $B^+\rightarrow \pi^+\chi_{c0}\rightarrow \pi^+K^+K^-$ & $(10^{-8})$ &$4.15_{-1.01}^{+0.89}(\omega_B)_{-0.60}^{+0.63}(a_2)_{-0.23}^{+0.24}(\Gamma_{\chi_{c0}})_{-0.02}^{+0.01}(m_0^{\pi})$&$-$\\
       $B^0\rightarrow \pi^0\chi_{c0}\rightarrow \pi^0\pi^+\pi^-$ & $(10^{-8})$ &$1.96_{-0.13}^{+0.32}(\omega_B)_{-0.28}^{+0.26}(a_2)_{-0.13}^{+0.16}(\Gamma_{\chi_{c0}})_{-0.00}^{+0.02}(m_0^{\pi})$&$-$\\
       $B^0\rightarrow \pi^0\chi_{c0}\rightarrow \pi^0 K^+K^-$ & $(10^{-8})$ &$2.06_{-0.36}^{+0.45}(\omega_B)_{-0.32}^{+0.30}(a_2)_{-0.10}^{+0.12}(\Gamma_{\chi_{c0}})_{-0.04}^{+0.01}(m_0^{\pi})$&$-$\\
       \hline \hline
\end{tabular}
\end{center}
\end{table}

We calculate the branching ratios for the decays of $B \rightarrow h\chi_{c0} \rightarrow h\pi^+\pi^-(K^+K^-)$ in Table~(\ref{1}), by using the differential branching ratios in Eq.~(\ref{eq:B}), and the decay amplitudes in the Appendix.
Compare our numerical results with current world average values from the PDG\cite{PDG2022} and the various theoretical predictions in PQCD, LCSR and QCDF in Table~(\ref{2}), and we do some analyses.

With a assumption that the reaction between the branching ratio of the quasi-two-body decay and the two-body framework satisfies
$\mathcal{B}(B^+\rightarrow h\chi_{c0}\rightarrow h\pi^+\pi^-)=\mathcal{B}(B^+\rightarrow h\chi_{c0})\cdot\mathcal{B}(\chi_{c0}\rightarrow \pi^+\pi^-)$, then we have PQCD prediction of branching ratio$\mathcal{B}(B^+\rightarrow K^+\chi_{c0})=\frac{\mathcal{B}(B^+\rightarrow K^+\chi_{c0}\rightarrow K^+\pi^+\pi^-)}{\mathcal{B}(\chi_{c0}\rightarrow \pi^+\pi^-)}=(1.42_{-0.92}^{+0.78})\times10^{-4}$, and $\mathcal{B}(B^+\rightarrow K^+\chi_{c0})=\frac{\mathcal{B}(B^+\rightarrow K^+\chi_{c0}\rightarrow K^+K^+K^-)}{\mathcal{B}(\chi_{c0}\rightarrow K^+K^-)}=(1.39_{-0.73}^{+0.54})\times 10^{-4}$ where the branching ratio of $\mathcal{B}(\chi_{c0}\rightarrow \pi^+\pi^-)=\frac{2}{3}\mathcal{B}(\chi_{c0}\rightarrow \pi\pi)=(5.67\pm0.22)\times10^{-3}$, $\mathcal{B}(\chi_{c0}\rightarrow K^+K^-)=(6.05\pm0.31)\times10^{-3}$\cite{PDG2022}.
The two results above predicted by PQCD agree well with the branching fractions $(1.51_{-0.13}^{+0.15})\times10^{-4}$ for the two-body decays $B^+\rightarrow K^+\chi_{c0}$ in the \textit{Review of Particle Physics}\cite{PDG2022}, respectively.
Our prediction for $\mathcal{B}(B^0\rightarrow K^0\chi_{c0})=(2.13_{-1.01}^{+1.54})\times 10^{-4}$ agree with data ($1.9\pm0.4)\times10^{-4}$ for two-body decays $B^0\rightarrow K^0\chi_{c0}$\cite{PDG2022}.

We contrast the various theoretical predictions for the  $B\rightarrow K\chi_{c0}$ cases of the investigated quasi-two-body and two-body decays.
The LCSR calculations mainly focus on $B^+\rightarrow K^+\chi_{c0}$ and the prediction value is $(1.0\pm0.6)\times10^{-4}$\cite{prd70074006}. Compared with previous PQCD calculations\cite{cpc44113104,prd71114008}, we update the charmonium distribution amplitudes and some of the input parameters in this study.
Our predictions are smaller than those of \cite{prd71114008} and closer to \cite{cpc44113104}.
The QCDF suffers endpoint divergences caused by spectator amplitudes and infrared divergences resulting from vertex diagrams.
The different treatment of these divergences as mentioned in the Introduction in \cite{0607221,plb619313,npb811155} lead to different numerical results.
Both our results in this work and the computations above are in excellent agreement with the available data for $B^+\rightarrow K^+\chi_{c0}$ and $B^0\rightarrow K^0\chi_{c0}$.

\begin{table}[!th]
\begin{center}
\caption{PQCD predictions of branching ratios for the two-body decays $B_{(s)}\rightarrow h\chi_{c0}[\chi_{c0} \rightarrow \pi^+\pi^-(K^+K^-)]$.}
\label{2}
\vspace{0cm}
\begin{tabular}{l c c c  c c c c } \hline \hline
Mode~~~~~&~~~~~Unit~~~~~&~~~~~This Work&~~~Data\cite{PDG2022}~~~&~~~PQCD~~~&~~~LCSR~~~&~~~QCDF~~~~~~\\
\hline

       $B^+\rightarrow K^+\chi_{c0}$ & $(10^{-4})$ & $1.42_{-0.92}^{+0.78}$ & $1.51_{-0.13}^{+0.15}$&$1.4^{+1.3}_{-0.9}$\cite{cpc44113104}&$1.0\pm0.6$\cite{prd70074006}&$1.05$\cite{0607221}\\
       &&&&5.61\cite{prd71114008}&&$0.78_{-0.35}^{+0.46}$\cite{plb619313}\\
       $B^0\rightarrow K^0\chi_{c0}$ & $(10^{-4})$ & $2.13_{-1.01}^{+1.54}$ & $1.9\pm0.4$&$1.3_{-0.8}^{+1.2}$\cite{cpc44113104}& -  &$1.13\sim5.19$\cite{npb811155} \\
       &&&&5.24\cite{prd71114008}&&\\
       $B_s^0\rightarrow \bar{K}^0\chi_{c0}$ & $(10^{-5})$ & $3.28_{-1.08}^{+1.51}$ &$-$ & $4.3_{-3.0}^{+4.4}$\cite{cpc44113104} & $-$&$-$\\
       $B^+\rightarrow \pi^+\chi_{c0}$ & $(10^{-5})$ & $0.69_{-0.26}^{+0.22}$ &$-$&$0.36_{-0.24}^{+0.37}$\cite{cpc44113104}&$-$&$-$ \\
       $B^0\rightarrow \pi^0\chi_{c0}$ & $(10^{-5})$ & $0.34_{-0.10}^{+0.13}$ &$-$&$-$&$-$&$-$ \\
       \hline \hline
\end{tabular}
\end{center}
\end{table}

Now, we turn our attention to $B\rightarrow h\chi_{c0}\rightarrow h\pi^+\pi^-(K^+K^-)$ with $h=\pi,\bar{K}^0$ decay models.
These decays, which proceed via a $b \rightarrow dc\bar{c}$ quark transition, are Cabibbo-suppressed decays.
Effects of $SU(3)$ breaking on distribution amplitudes makes a negative contribution to decay, causing the branching ratio to be small.
Experimentally, only the BaBar collaboration reported the upper bound $0.1\times 10^{-6}$ on the branching ratio for $B^+\rightarrow \pi^+\chi_{c0}\rightarrow \pi^+\pi^+\pi^-$\cite{prd79072006}.
Our result is $3.93_{-1.46}^{+1.69}\times10^{-8}$, which is in consistent with the scope of the measured data by BaBar.
The data for decay modes $B^+\rightarrow \pi^+\chi_{c0}\rightarrow \pi^+K^+K^-$, $\sqrt{2}B^0\rightarrow\pi^0\chi_{c0}\rightarrow \pi^0\pi^+\pi^-$ and $\sqrt{2}B^0\rightarrow \pi^0\chi_{c0}\rightarrow\pi^0 K^+K^-$ are around $10^{-8}$, which can be examined in the forthcoming experiments.
Since these Cabibbo-suppressed decays are still received less attention in other approaches, we are waiting for future comparison.

For the quasi-two-body processes $B^+\rightarrow \pi^+\chi_{c0}\rightarrow \pi^+\pi^+\pi^-$ and $B^+\rightarrow K^+\chi_{c0}\rightarrow K^+\pi^+\pi^-$, which have an identical step $\chi_{c0}\rightarrow \pi^+\pi^-$, the difference of these two decay modes originated from the bachelor particles pion and kaon. Assuming factorization and flavor-$SU(3)$ symmetry, the ratio $R_{\chi_{c0}}$ for the branching fractions of these two processes is
\begin{eqnarray}
R_{\chi_{c0}}=\frac{\mathcal{B}(B^+\rightarrow \pi^+\chi_{c0}\rightarrow\pi^+\pi^+\pi^-)}{\mathcal{B}(B^+\rightarrow K^+\chi_{c0}\rightarrow K^+\pi^+\pi^-)}\approx \mid\frac{V_{cd}}{V_{cs}}\mid^2 \cdot \frac{f_{\pi}^2}{f_k^2}.
\label{eq:R1}
\end{eqnarray}
With the result
\begin{eqnarray}
 \mid \frac{V_{cd}}{V_{cs}}\mid \cdot \frac{f_{\pi}}{f_k}=0.189,
\label{eq:r1}
\end{eqnarray}
in \textit{Review of Particle Physics} \cite{PDG2022}, one has $R_{\chi_{c0}}\approx 0.036$.
It still fits expectations from our PQCD anticipated ratio
\begin{eqnarray}
R_{\chi_{c0}}=\frac{\mathcal{B}(B^+\rightarrow \pi^+\chi_{c0}\rightarrow\pi^+\pi^+\pi^-)}{\mathcal{B}(B^+\rightarrow K^+\chi_{c0}\rightarrow K^+\pi^+\pi^-)}=0.049_{-0.009}^{+0.020}.
\label{eq:R2}
\end{eqnarray}

%%%%%%%%%%%%%%%%%%%%%%%%%%%%%%%%%%%%%%%%%%%%%%
\begin{figure}[!th]
\begin{center}
\includegraphics[width=0.5\textwidth]{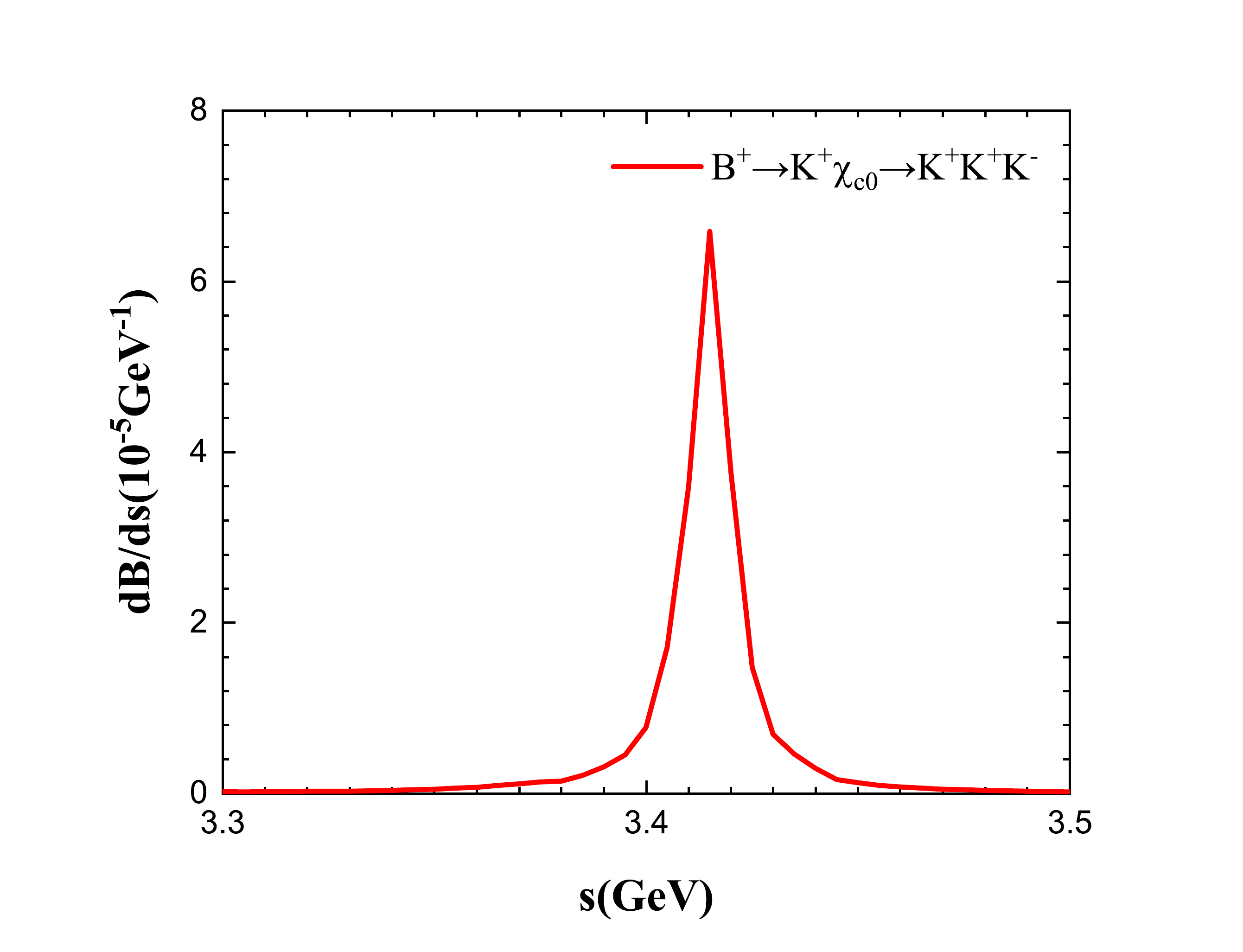}
\caption{The $m_{KK}$ dependence of decay rates $dB/dm_{KK}$ for the $B \rightarrow K\chi_{c0} \rightarrow KKK$ .}
\label{fig:fig3}
\end{center}
\end{figure}

In Fig.~\ref{fig:fig3}, we show the distribution of branching ratios for decays modes $B^+\rightarrow K^+\chi_{c0} \rightarrow K^+K^+K^-$.
The mass of $\chi_{c0}$ is visible as a narrow peaks near $3.414\,{\rm GeV}$.
We find that the central portion of the branching ratios lies in the region around the pole mass of the $\chi_{c0}$ resonance as shown by the distribution of the branching ratios in the $\pi\pi$ invariant mass.

\section{CONCLUSION}
We studied the nonfactorizable contributions to these factorization-forbidden quasi-two-body decays $B \to K\chi_{c0} \to K\pi\pi(KK)$, $B_s \to \bar{K}^0\chi_{c0} \to \bar{K}^0\pi\pi(KK)$, and $B_s \to \pi\chi_{c0} \to \pi\pi\pi(KK)$ in PQCD approach in this work.
Our predictions for the branching ratios are summarized in Table~\ref{1} and compared with other theoretical results.
The obtained branching ratios of $B\rightarrow K\chi_{c0}$ decay are essentially consistent with the current data.
For the decay involving $\pi$ or $\bar{K}$ in the final state not yet measured, the calculated branching ratios will be further tested by experiments in the near future.
By utilizing the flavor-$SU(3)$ symmetry to examine quasi-two-body decays with the same intermediate step, we were able to establish the ratio $R_{\chi_{c0}}$ for processes $B^+\rightarrow \pi^+\chi_{c0}\rightarrow \pi^+\pi^+\pi^-$ and $B^+\rightarrow K^+\chi_{c0}\rightarrow K^+\pi^+\pi^-$.
The ratio $R_{\chi_{c0}}$ is predicted by PQCD to be $0.049$, which is close to the value $0.036$ reported in Review of Particle Physics.
We also display the distribution of branching ratios for various decay modes in invariant mass, and we discover that the majority of the branching ratios are located in the vicinity of the $\chi_{c0}$ resonance's pole mass.

\section{ACKNOWLEDGEMENTS}
Many thanks to Wen-Fei Wang, Da-Cheng Yan and Jun Hua for valuable discussions.

\begin{appendix}
\section{Decay amplitudes}
The concerned quasi-two-body decay amplitudes are given in the PQCD approach by
\begin{eqnarray}
\mathcal{A}\left(B^+\rightarrow\pi^+[\chi_{c0}\rightarrow]\pi^+\pi^-\right)&=&\frac{G_F}{\sqrt{2}}\left\{V_{cb}^*V_{cd} c_2
          M_{e\pi}^{LL}-V_{tb}^*V_{td}\left[(c_4+c_{10})M_{e\pi}^{LL}+(c_6+c_8)M_{e\pi}^{SP}\right]\right\},   \label{eq:A1} \\
\mathcal{A}\left(B^+\rightarrow K^+[\chi_{c0}\rightarrow]\pi^+\pi^-\right)&=&\frac{G_F}{\sqrt{2}}\left\{V_{cb}^*V_{cs} c_2
          M_{eK}^{LL}-V_{tb}^*V_{ts}\left[(c_4+c_{10})M_{eK}^{LL}+(c_6+c_8)M_{eK}^{SP}\right]\right\},       \label{eq:A2} \\
\mathcal{A}\left(B^0\rightarrow\pi^0[\chi_{c0}\rightarrow]\pi^+\pi^-\right)&=&\frac{G_F}{\sqrt{2}}\left\{V_{cb}^*V_{cd} c_2
          M_{e\pi}^{LL}-V_{tb}^*V_{td}\left[(c_4+c_{10})M_{e\pi}^{LL}+(c_6+c_8)M_{e\pi}^{SP}\right]\right\},    \label{eq:A3} \\
\mathcal{A}\left(B^0\rightarrow K^0[\chi_{c0}\rightarrow]\pi^+\pi^-\right)&=&\frac{G_F}{\sqrt{2}}\left\{V_{cb}^*V_{cs} c_2
         M_{eK}^{LL} -V_{tb}^*V_{ts}\left[(c_4+c_{10})M_{eK}^{LL}+(c_6+c_8)M_{eK}^{SP}\right]\right\},        \label{eq:A4} \\
\mathcal{A}\left(B_s^0\rightarrow \bar{K}^0[\chi_{c0}\rightarrow]\pi^+\pi^-\right)&=&\frac{G_F}{\sqrt{2}}\left\{V_{cb}^*V_{cd} c_2
         M_{eK}^{LL} -V_{tb}^*V_{td}\left[(c_4+c_{10})M_{eK}^{LL}+(c_6+c_8)M_{eK}^{SP}\right]\right\},         \label{eq:A5}
\end{eqnarray}
where $G_F$ is the Fermi coupling constant, $V^{,}\textmd{s}$ are the Cabibbo-Kobayashi-Maskawa matrix elements,
and $c_i$ is Wilson coefficients. The amplitudes appeared in above equations are written as
\begin{eqnarray}
M_{eK(\pi)}^{LL}&=&-16\sqrt{\frac{2}{3}}\pi C_F m_B^4\int_0^1 dx_B dz dx_3\int_0^\infty b_B db_B b_3db_3\phi_B(x_B,b_B) \nonumber\\
&&\{[(\eta-1)(\sqrt{\eta}r\phi_{\pi\pi}^s(z)-(\eta+1)(x_B+z-1)\phi_{\pi\pi}^v(z))\phi^A(x_3)\nonumber\\
&&+r_3(4\sqrt{\eta}r\phi_{\pi\pi}^s(z)+(x_3-\eta(x_3+x_B+2z-2))\phi_{\pi\pi}^v(z))\phi^P(x_3)\nonumber\\
&&+r_3((\eta-1)x_3-\eta x_B)\phi_{\pi\pi}^v(z)\phi^T(x_3)]E_a(t_a)h_a(x_B,z,x_3;b_B,b_3)\nonumber\\
&&+[(\eta-1)(\sqrt{\eta}r\phi_{\pi\pi}^s(z)+((\eta-1)x_3+x_B-(\eta+1)z)\phi_{\pi\pi}^v(z))\phi^A(x_3)\nonumber\\
&&+r_3(4\sqrt{\eta}r\phi_{\pi\pi}^s(z)+(\eta(x_3+x_B-2z)-x_3)\phi_{\pi\pi}^v(z))\phi^P(x_3)\nonumber\\
&&+r_3((\eta-1)x_3-\eta x_B)\phi_{\pi\pi}^v(z)\phi^T(x_3)]E_b(t_b)h_b(x_B,z,x_3;b_B,b_3)\}
\label{eq:A6}
\end{eqnarray}
\begin{eqnarray}
M_{eK(\pi)}^{SP}&=&-16\sqrt{\frac{2}{3}}\pi C_F m_B^4\int_0^1 dx_B dz dx_3\int_0^\infty b_B db_B b_3db_3\phi_B(x_B,b_B) \nonumber\\
&&\{[(\eta-1)(\sqrt{\eta}r\phi_{\pi\pi}^s(z)-((\eta-1)x_3+x_B+(\eta+1)(z-1))\phi_{\pi\pi}^v(z))\phi^A(x_3)\nonumber\\
&&+r_3(4\sqrt{\eta}r\phi_{\pi\pi}^s(z)+(x_3-\eta(x_3+x_B+2z-2))\phi_{\pi\pi}^v(z))\phi^P(x_3)\nonumber\\
&&-r_3((\eta-1)x_3-\eta x_B)\phi_{\pi\pi}^v(z)\phi^T(x_3)]E_a(t_a)h_a(x_B,z,x_3;b_B,b_3)\nonumber\\
&&+[(\eta-1)(\sqrt{\eta}r\phi_{\pi\pi}^s(z)+(\eta+1)(x_B-z)\phi_{\pi\pi}^v(z))\phi^A(x_3)\nonumber\\
&&+r_3(4\sqrt{\eta}r\phi_{\pi\pi}^s(z)+(\eta(x_3+x_B-2z)-x_3)\phi_{\pi\pi}^v(z))\phi^P(x_3)\nonumber\\
&&-r_3((\eta-1)x_3-\eta x_B)\phi_{\pi\pi}^v(z)\phi^T(x_3)]E_b(t_b)h_b(x_B,z,x_3;b_B,b_3)\},
\label{eq:A7}
\end{eqnarray}
with the $r_c=m_c/m_B$ and $r_3=m_{0}^{h}/m_B$.
The evolution factors in above formulas are given by
\begin{eqnarray}
E_{a(b)}(t)=\alpha_s(t)\exp[-S_{ab}(t)].
\label{eq:A8}
\end{eqnarray}
The hard functions $h_{a(b)}$, the hard scales $t_{a(b)}$, and factor $S_{ab}(t)$ have their explicit expressions in the Appendix of \cite{prd78014018}.
\end{appendix}

%%------------------------------------%%%%%%%%%%%%%%%

\end{document}